\RequirePackage{lineno}

\documentclass[aps,prc,twocolumn,groupedaddress,showpacs,preprintnumbers,superscriptaddress,floatfix,amsmath,amssymb]{revtex4}
\usepackage [dvips]{graphics, color,epsfig}
\usepackage{multirow}
\usepackage{xspace}
\usepackage{dcolumn}
\usepackage{bm}
\usepackage{hyperref}
\usepackage{hyperref}

\makeatletter{}
\newcommand{\efficiencystatement}{The systematic uncertainty due to the uncertainty on the associated particle's reconstruction efficiency (5\%) and background level extraction (2\%) are not shown.\xspace}
\newcommand{\sNNtwohundred}{$\sqrt{s_{{NN}}}$ = 200 GeV\xspace}
\newcommand{\sqrts}{$\sqrt{s}$\xspace}

\newcommand{\stdassoc}{$1.5$~GeV/$c$ $<$ $p_T^{\mathrm{associated}}$ $<$ $p_T^{\mathrm{trigger}}$\xspace}
\newcommand{\stdtrig}{$3$~$<$ $p_T^{\mathrm{trigger}}$ $<$ $6$~GeV/$c$\xspace}
\newcommand{\trigrange}[2]{#1 $< p_T^{\mathrm{trigger}} <$ #2~GeV/$c$\xspace}

\newcommand{\assocrange}[2]{#1 $< p_T^{\mathrm{associated}} < $ #2~GeV/$c$\xspace}

\newcommand{\pttrig}{$p_T^{\mathrm{trigger}}$\xspace}
\newcommand{\ptassoc}{$p_T^{\mathrm{associated}}$\xspace}
\newcommand{\npart}{$N_{\mathrm{part}}$\xspace}

\newcommand{\pT}{$p_T$\xspace}
\newcommand{\highpT}{high-\pT}

\newcommand{\pp}{$p$+$p$\xspace}
\newcommand{\Cu}{Cu+Cu\xspace}
\newcommand{\Au}{Au+Au\xspace}

\newcommand{\dAu}{$d$+Au\xspace}
\newcommand{\AplusA}{$A$+$A$\xspace}

\newcommand{\GeV}{GeV/$c$\xspace}
\newcommand{\dphi}{$\Delta\phi$\xspace}
\newcommand{\deta}{$\Delta\eta$\xspace}
\newcommand{\vtwo}{$v_2$\xspace}
\newcommand{\vthree}{$v_{3}$\xspace}

\newcommand{\ns}{near-side\xspace}

\newcommand{\jl}{jet-like\xspace}
\newcommand{\PT}{PYTHIA\xspace}
\newcommand{\lam}{$\Lambda$\xspace}
\newcommand{\alam}{$\bar{\Lambda}$\xspace}
\newcommand{\lamalamalt}{$\Lambda$, $\bar{\Lambda}$\xspace}

\newcommand{\vzero}{$V^{0}$\xspace}
\newcommand{\vzeroh}{$V^{0}$-h\xspace}
\newcommand{\hvzero}{h-$V^{0}$\xspace}
\newcommand{\kaon}{$K^0_S$\xspace}
\newcommand{\hkaon}{h-$K^0_S$\xspace}
\newcommand{\hlam}{h-$\Lambda$\xspace}
\newcommand{\kaonh}{$K^0_S$-h\xspace}
\newcommand{\lamh}{$\Lambda$-h\xspace}
\newcommand{\hh}{h-h\xspace}

\newcommand{\ntrig}{$N_{\mathrm{trigger}}$\xspace}

\newcommand{\eref}[1]{Equation~\ref{#1}}
\newcommand{\Sref}[1]{Section~\ref{#1}}
\newcommand{\Fref}[1]{Fig.~\ref{#1}}
\newcommand{\Tref}[1]{Tab.~\ref{#1}}

\newcommand{\roughly}{$\approx$\xspace}

\newcommand{\jlc}{jet-like correlation\xspace}

\newcommand{\jly}{jet-like yield\xspace}
\newcommand{\jlys}{jet-like yields\xspace}

\newcommand{\dsqnwitharg}{\frac{d^2N}{d\Delta\phi\, d\Delta\eta}(\Delta\phi,\Delta\eta)}

\newcommand{\dNdEta}{\frac{dN}{d\Delta\eta}}

\def\Deta{\mbox{$\Delta\eta$}}
\def\Dphi{\mbox{$\Delta\phi$}}
\newcommand{\detano}{\ensuremath{\Delta\eta}}

\newcommand{\bEta}{b_{\Delta\eta}}

\newcommand{\Yjeta}{Y^{\Delta\eta}_{J}}

\begin{document}
 
\title{Near-side azimuthal and pseudorapidity correlations using neutral strange baryons and mesons in \dAu, \Cu and \Au  
collisions at $\sqrt{s_{NN}}$~=~200~GeV} 

\makeatletter{}\affiliation{AGH University of Science and Technology, FPACS, Cracow 30-059, Poland}
\affiliation{Argonne National Laboratory, Argonne, Illinois 60439}
\affiliation{Brookhaven National Laboratory, Upton, New York 11973}
\affiliation{University of California, Berkeley, California 94720}
\affiliation{University of California, Davis, California 95616}
\affiliation{University of California, Los Angeles, California 90095}
\affiliation{Central China Normal University, Wuhan, Hubei 430079}
\affiliation{University of Illinois at Chicago, Chicago, Illinois 60607}
\affiliation{Creighton University, Omaha, Nebraska 68178}
\affiliation{Czech Technical University in Prague, FNSPE, Prague, 115 19, Czech Republic}
\affiliation{Nuclear Physics Institute AS CR, 250 68 Prague, Czech Republic}
\affiliation{Frankfurt Institute for Advanced Studies FIAS, Frankfurt 60438, Germany}
\affiliation{Institute of Physics, Bhubaneswar 751005, India}
\affiliation{Indian Institute of Technology, Mumbai 400076, India}
\affiliation{Indiana University, Bloomington, Indiana 47408}
\affiliation{Alikhanov Institute for Theoretical and Experimental Physics, Moscow 117218, Russia}
\affiliation{University of Jammu, Jammu 180001, India}
\affiliation{Joint Institute for Nuclear Research, Dubna, 141 980, Russia}
\affiliation{Kent State University, Kent, Ohio 44242}
\affiliation{University of Kentucky, Lexington, Kentucky, 40506-0055}
\affiliation{Korea Institute of Science and Technology Information, Daejeon 305-701, Korea}
\affiliation{Institute of Modern Physics, Chinese Academy of Sciences, Lanzhou, Gansu 730000}
\affiliation{Lawrence Berkeley National Laboratory, Berkeley, California 94720}
\affiliation{Max-Planck-Institut fur Physik, Munich 80805, Germany}
\affiliation{Michigan State University, East Lansing, Michigan 48824}
\affiliation{National Research Nuclear Univeristy MEPhI, Moscow 115409, Russia}
\affiliation{National Institute of Science Education and Research, Bhubaneswar 751005, India}
\affiliation{National Cheng Kung University, Tainan 70101 }
\affiliation{Ohio State University, Columbus, Ohio 43210}
\affiliation{Institute of Nuclear Physics PAN, Cracow 31-342, Poland}
\affiliation{Panjab University, Chandigarh 160014, India}
\affiliation{Pennsylvania State University, University Park, Pennsylvania 16802}
\affiliation{Institute of High Energy Physics, Protvino 142281, Russia}
\affiliation{Purdue University, West Lafayette, Indiana 47907}
\affiliation{Pusan National University, Pusan 46241, Korea}
\affiliation{Rice University, Houston, Texas 77251}
\affiliation{University of Science and Technology of China, Hefei, Anhui 230026}
\affiliation{Shandong University, Jinan, Shandong 250100}
\affiliation{Shanghai Institute of Applied Physics, Chinese Academy of Sciences, Shanghai 201800}
\affiliation{State University Of New York, Stony Brook, NY 11794}
\affiliation{Temple University, Philadelphia, Pennsylvania 19122}
\affiliation{Texas A\&M University, College Station, Texas 77843}
\affiliation{University of Birmingham, Birmingham, United Kingdom}
\affiliation{University of Texas, Austin, Texas 78712}
\affiliation{University of Houston, Houston, Texas 77204}
\affiliation{Tsinghua University, Beijing 100084}
\affiliation{United States Naval Academy, Annapolis, Maryland, 21402}
\affiliation{Valparaiso University, Valparaiso, Indiana 46383}
\affiliation{Variable Energy Cyclotron Centre, Kolkata 700064, India}
\affiliation{Warsaw University of Technology, Warsaw 00-661, Poland}
\affiliation{Wayne State University, Detroit, Michigan 48201}
\affiliation{World Laboratory for Cosmology and Particle Physics (WLCAPP), Cairo 11571, Egypt}
\affiliation{Yale University, New Haven, Connecticut 06520}

\author{B.~Abelev}\affiliation{Yale University, New Haven, Connecticut 06520, USA}
\author{L.~Adamczyk}\affiliation{AGH University of Science and Technology, FPACS, Cracow 30-059, Poland}
\author{J.~K.~Adkins}\affiliation{University of Kentucky, Lexington, Kentucky, 40506-0055}
\author{G.~Agakishiev}\affiliation{Joint Institute for Nuclear Research, Dubna, 141 980, Russia}
\author{M.~M.~Aggarwal}\affiliation{Panjab University, Chandigarh 160014, India}
\author{Z.~Ahammed}\affiliation{Variable Energy Cyclotron Centre, Kolkata 700064, India}
\author{I.~Alekseev}\affiliation{Alikhanov Institute for Theoretical and Experimental Physics, Moscow 117218, Russia}
\author{A.~Aparin}\affiliation{Joint Institute for Nuclear Research, Dubna, 141 980, Russia}
\author{D.~Arkhipkin}\affiliation{Brookhaven National Laboratory, Upton, New York 11973}
\author{E.~C.~Aschenauer}\affiliation{Brookhaven National Laboratory, Upton, New York 11973}
% ...

\collaboration{STAR Collaboration}\noaffiliation 
 
\date{\today}             
\begin{abstract}
We present measurements of the \ns of triggered di-hadron correlations using neutral strange baryons (\lam, \alam) 
and mesons (\kaon) at intermediate transverse momentum (3~$<$~\pT~$<$~6~\GeV) to look for possible flavor and baryon/meson dependence.  This study is performed in \dAu, \Cu and \Au collisions 
at \sNNtwohundred measured by the STAR experiment at RHIC.
The \ns di-hadron correlation contains two structures, a peak which is narrow in azimuth and pseudorapidity consistent with correlations due to jet fragmentation, and a correlation in azimuth which is broad in pseudorapidity.  The particle composition of the \jlc is determined using identified associated particles.  The dependence of the conditional yield of the \jlc on the trigger particle momentum, associated particle momentum, and centrality for correlations with unidentified trigger particles are presented.  
The neutral strange particle composition in jet-like correlations with unidentified charged particle triggers is not well described by \PT.  However, the yield of unidentified particles in jet-like correlations with neutral strange particle triggers is described reasonably well by the same model.

\end{abstract}

\maketitle

\section{Introduction}\label{section-introduction}
Ultrarelativistic heavy-ion collisions create a unique environment  for the investigation 
of nuclear matter at extreme temperatures and energy densities. 
Measurements of nuclear modification factors~\cite{Adler:2003kg,Adams:2003am,CMS:2012aa,Aamodt:2010jd,Aad:2015wga} 
show that the nuclear medium created is nearly opaque to partons with large transverse momentum (\pT). Anisotropic flow measurements demonstrate  that the medium exhibits partonic degrees of freedom and has properties close to those expected of a perfect fluid~\cite{Adams:2003am,Adams:2004bi,Adare:2006ti,Alver:2006wh}.

Studies of jets in heavy ion collisions are possible through single particle measurements~\cite{Adler:2003kg,Adams:2003am,CMS:2012aa,Aamodt:2010jd}, di-hadron correlations~\cite{Agakishiev:2011st,Adams:2005ph,STARConical:2008nd,Adler:2002tq,Adams:2006yt,RidgePaper:2009qa,Abelev:2009jv,Agakishiev:2014ada,Aamodt:2011vg,Chatrchyan:2011eka,Chatrchyan:2012wg}, and measurements of reconstructed jets~\cite{Aad:2012vca,Aad:2010bu,CMS:2012aa,Chatrchyan:2013kwa,Abelev:2012ej} and their correlations with hadrons~\cite{Adamczyk:2013jei,Adam:2015doa}.  Measurements of reconstructed jets provide direct evidence for partonic energy loss in the medium.  Di-hadron and jet-hadron correlations enable studies at intermediate momenta, where the interplay between jets and the medium is important and direct jet reconstruction is challenging.

Properties of jets have been studied extensively using di-hadron correlations relative to a trigger particle with large transverse momentum at the Relativistic Heavy Ion Collider (RHIC)~\cite{Agakishiev:2011st,Adams:2005ph,STARConical:2008nd,Adler:2002tq,Adams:2006yt,RidgePaper:2009qa,Abelev:2009jv,Agakishiev:2014ada} and the Large Hadron Collider (LHC)~\cite{Aamodt:2011vg,Chatrchyan:2011eka,Chatrchyan:2012wg}.
Systematic studies of associated particle distributions on the opposite side of the trigger particle in azimuth (\dphi~$\approx$~180$^{\circ}$) revealed significant modification, including the disappearance of the peak at intermediate transverse momentum, approximately 2--4 \GeV~\cite{Adler:2002tq,Adare:2007vu} and its reappearance at high \pT~\cite{Adams:2006yt,Adare:2010ry}. 
The associated particle distribution on the near side of the trigger particle, the subject of this paper, is also significantly modified in central \Au collisions~\cite{Adams:2005ph,RidgePaper:2009qa,Adler:2005ee}.   In \pp and \dAu collisions, there is a peak that is narrow in azimuth and pseudorapidity (\deta) around the trigger particle, which we refer to as the \jlc.  In \Cu and \Au collisions this peak is observed to be broader than that in \dAu collisions, although the yields are comparable~\cite{Agakishiev:2011st}.  
Besides the shape modifications of jet-like correlations at intermediate transverse momenta, the production mechanism of hadrons may differ from simple fragmentation.
In central \AplusA collisions baryon production is enhanced relative to that in \pp collisions~\cite{Abelev:2006jr,Agakishiev:2011ar,Abelev:2013xaa}.
The baryon to meson ratios measured in \Au collisions increase with increasing $p_T$ until reaching a maximum of approximately three times that observed in \pp collisions at $p_T\approx$~3~GeV/$c$ in both the strange and non-strange quark sectors. A fall-off of the baryon to meson ratio is observed for $p_T>$~3~GeV/$c$ and
both the strange and non-strange baryon to meson ratios in \Au collisions approach the values measured in \pp collisions at $p_T\approx$~6~GeV/$c$.
 Using statistical separation di-hadron correlation studies with pion and non-pion triggers~\cite{Abdelwahab:2014cvd}  showed that significant enhancement of near-side jet-like yields in central Au+Au collisions relative to \dAu collisions is present for pion triggered correlations. In contrast, for the non-pion triggered sample which consists mainly of protons and charged kaons no statistically significant difference is observed.

In this paper, studies of two-particle correlations on the near-side in \dAu, \Cu and \Au collisions at \sNNtwohundred measured by the STAR experiment are presented. Results from two-particle correlations in pseudorapidity and azimuth for neutral strange baryons (\lamalamalt) and mesons (\kaon) at intermediate \pT  (3~$<$~\pT$~<$~6 \GeV) in the different collision systems are compared to unidentified charged particle correlations (\hh). 
Both identified strange trigger particles associated with unidentified charged particles (\kaonh, \lamh) and unidentified charged trigger particles associated with identified strange particles (\hkaon, \hlam) are studied. 
The \ns \jl yield is studied as a function of centrality  of the collision and transverse momentum of trigger and associated particles to look for possible flavor and baryon/meson dependence.  The composition of the \jlc is studied using identified associated particles to investigate possible medium effects on particle production.  The results are compared to expectations from \PT~\cite{Sjostrand:2006za}.

\section{Experimental setup and particle reconstruction}\label{section-experiment}
The Solenoidal Tracker At RHIC (STAR) experiment~\cite{Ackermann:2002ad} is a multipurpose spectrometer with a full azimuthal coverage consisting of several detectors inside a large solenoidal magnet with a uniform  magnetic field of 0.5~T applied parallel to the beam line. This analysis is based exclusively on charged particle tracks detected and reconstructed in the Time Projection Chamber (TPC)~\cite{Anderson:2003ur} with a pseudorapidity acceptance $|\eta|<$~1.5. The TPC has in total 45 pad rows in the radial direction allowing up to 45 independent spatial and energy loss ($dE/dx$) measurements for each charged particle track. 
Charged particle tracks used in this  analysis were required to have at least 15 fit points in the TPC, a distance of closest approach to the primary vertex of less than 1 cm and a pseudorapidity $|\eta|<$~1.0.  These tracks are referred to as charged hadron tracks because the majority of them come from charged hadrons. 
The results presented in this paper are based on analysis of data from \dAu, \Cu, and \Au collisions at \sNNtwohundred taken by the STAR experiment 
in 2003, 2005, and 2004, respectively. 

For \dAu collisions, the events analyzed were selected using a minimally biased (MB) trigger requiring 
at least one beam-rapidity neutron in the Zero Degree Calorimeter (ZDC), located 18~m from the nominal interaction point in the Au beam direction 
and accepting 95$\pm$3\% of the hadronic cross section~\cite{Adams:2003im}.  For \Cu collisions, the MB trigger was based on the combined 
signals from the Beam-Beam Counters (BBC) placed at forward pseudorapidity (3.3~$<|\eta|<$~5.0) and a coincidence  between the two ZDCs.  
The MB \Au events 
required a coincidence between the two ZDCs, a signal in both BBCs and a minimum charged particle multiplicity in an array of scintillator slats aligned parallel to the beam axis and 
arranged in a barrel, the Central Trigger Barrel (CTB), to reject non-hadronic interactions.  
An additional online trigger for central \Au collisions was used to sample the most central 12\% of the total hadronic cross section. This trigger was based on the energy deposited in the ZDCs in combination with the multiplicity in the CTB. Centrality selection is based on the primary charged particle multiplicity $N_{ch}$ within the pseudorapidity range $|\eta|<$~0.5, as in~\cite{Ackermann:2000tr,Adler:2002xw}.  Calculation of the number of participating nucleons, \npart, in each centrality class is done as in~\cite{Adamczyk:2013tvk,Abelev:2008ab,Abelev:2008zk}.

In order to achieve a more uniform detector acceptance in \Cu and \Au data sets, only those events with a primary collision vertex position along the beam axis ($z$) within 30~cm of the center of the STAR detector were used for the analysis.  For \dAu collisions this vertex position selection was extended to $|z|~<$~50~cm.  
The number of events after the vertex cuts in individual data samples is summarized in~\Tref{table-eventcount}.

\begin{table}[b!]
\caption{Number of events after cuts (see text) in the data samples analyzed.}
\label{tabevents}
\begin{tabular}{c c c}\hline \hline
System  & Centrality &  No. of events [10$^6$]\\ \hline \hline
\dAu    &  0-95\%   &  3\\ 
\Cu   &  0-60\% &  38\\ 
\Au   &  0-80\% &  28\\ 
\Au   &  0-12\% &  17\\ \hline \hline
\end{tabular}\label{table-eventcount}
\end{table}

We identify weakly decaying neutral strange (\vzero) particles  \lam, \alam and \kaon by topological 
reconstruction of their decay vertices from their charged hadron daughters  measured in the TPC as described in~\cite{Adler:2002uv}:
\begin{eqnarray}\label{vzerodecays}
\Lambda &\rightarrow& p+\pi^-, BR=(63.9\pm0.5)\% \nonumber \\
\bar{\Lambda} &\rightarrow& \bar{p}+\pi^+,  BR=(63.9\pm0.5)\% \\ 
K^{0}_{S} &\rightarrow& \pi^+ + \pi^-,  BR=(68.95\pm0.14)\% \nonumber
\end{eqnarray}
 where $BR$ denotes the branching ratio.
The \vzero reconstruction software pairs oppositely charged particle tracks into \vzero candidates. Reconstructed \lam and \kaon particles are required to be within $|\eta|<$~1.0. Topological cuts are optimized for each data set and chosen to have a signal-to-background ratio of at least 15:1.  For the analyses presented here, no difference was observed between results with \lam and \alam trigger particles.  Therefore the correlations with \lam and \alam trigger particles were combined to increase the statistical significance of the results.  In the remainder of the discussion the combined particles are refered to simply as \lam baryons.

\section{Method}\label{section-method}

\subsection{Correlation technique}
The analysis in this paper follows the method in~\cite{Agakishiev:2011st}.  A \highpT trigger particle was selected and the raw distribution of associated tracks relative to that trigger particle in pseudorapidity (\deta) and azimuth (\dphi) is formed.  This distribution, ${d^2N_{\mathrm{raw}}}/{d\Delta\phi\, d\Delta\eta}$, is normalized by the number of trigger particles, \ntrig, and corrected for the efficiency and acceptance of associated tracks:
\begin{eqnarray}\label{eq:di-hadronDist}
\dsqnwitharg & = & \frac{1}{N_{\mathrm{trigger}}} \frac{d^2N_{\mathrm{raw}}}{d\Delta\phi d\Delta\eta} \nonumber \\
& & \frac{1}{\varepsilon_{\mathrm{assoc}}(\phi,\eta)} \frac{1}{\varepsilon_{\mathrm{pair}}(\Delta\phi,\Delta\eta)}.
\end{eqnarray}
The efficiency correction $\varepsilon_{\mathrm{assoc}}(\phi,\eta)$ is a correction for the single particle reconstruction efficiency in TPC and $\varepsilon_{\mathrm{pair}}(\Delta\phi,\Delta\eta)$ is a correction for the finite TPC track-pair acceptance in \dphi and \deta, including track merging effects.  Since the correlations are normalized by the number of trigger particles, the efficiency correction is only applied for the associated particle.
The fully corrected correlation functions are averaged between positive and negative \dphi and \deta regions and are reflected about \dphi~=~0 and \deta~=~0 in the plots.

\subsection{Single particle efficiency correction}
For unidentified charged associated particles, the efficiency correction  $\varepsilon_{\mathrm{assoc}}(\phi,\eta)$ is the correction for charged particles, identical to that applied in~\cite{Agakishiev:2011st}.  This single charged track reconstruction efficiency  is determined as a function of \pT, $\eta$, and centrality by simulating the TPC response to a particle and embedding the simulated signals into a real event.  The efficiency is found to be approximately constant for \pT $>$~2~\GeV and ranges from around 75\% for central \Au events to around 85\% for peripheral \Cu events.  The efficiency for reconstructing a track in \dAu events is 89\%.  

For identified associated strange particles, the reconstruction efficiency  $\varepsilon_{\mathrm{assoc}}(\phi,\eta)$  is determined in a similar way, but forcing the simulated particle to decay through the channel in \eref{vzerodecays} and then correcting for the respective branching ratio.  The efficiency for reconstructing \lam, \alam, and \kaon ranges from 8\% to 15\%, increasing with momentum and decreasing with system size~\cite{Aggarwal:2010ig}. 
No correction for the reconstruction efficiency is applied for identified trigger particles because the reconstruction efficiency does not vary significantly within the \pttrig bins used in this analysis and the correlation function is normalized by the number of trigger particles.

The systematic uncertainty associated with the efficiency correction for unidentified associated particles is 5\% and is strongly correlated across centralities and \pT bins within each data set but not between data sets. For identified associated particle ratios the systematic uncertainties on the efficiency correction partially cancel out and are negligible compared to the statistical uncertainties.

For the inclusive spectra the feeddown correction due to secondary \lam baryons from $\Xi$ baryon decays is 15\%, independent of \pT~\cite{Agakishiev:2011ar}.  For identified \lam trigger particles, we assume that feeddown lambdas do not change the correlation.  Correlations with $\Xi$ triggers were performed to check this assumption.  
For identified associated particles, we assume the same correlation between primary and secondary \lam particles and correct the yield of \lam associated particles by reducing the yield by 15\%.

\subsection{Pair acceptance correction}\label{acceptanceCorrection}
The requirement that each track falls within $|\eta|<1.0$ in TPC results in a limited acceptance for track pairs.  The geometric acceptance for a track pair is \roughly 100\%  for \deta \roughly~$0$ and close to 0\% near \deta \roughly~$2$.  The track pair acceptance is limited in azimuth by the 12 TPC sector boundaries, leading to dips in the acceptance of track pairs in \dphi.  To correct for the limited geometric acceptance, a mixed event analysis was performed using trigger particles from one event combined with associated particles from another event, as done in~\cite{RidgePaper:2009qa}.  The event vertices were required to be within 2 cm of each other along the beam axis and the events were required to have the same charged particle multiplicity within 10 particles.  To increase statistics of the mixed event sample, each event with a trigger particle was mixed with ten other events.

\subsection{Yield extraction}\label{yieldmethod}
\begin{figure}
\rotatebox{0}{\resizebox{8.5cm}{!}{
\includegraphics{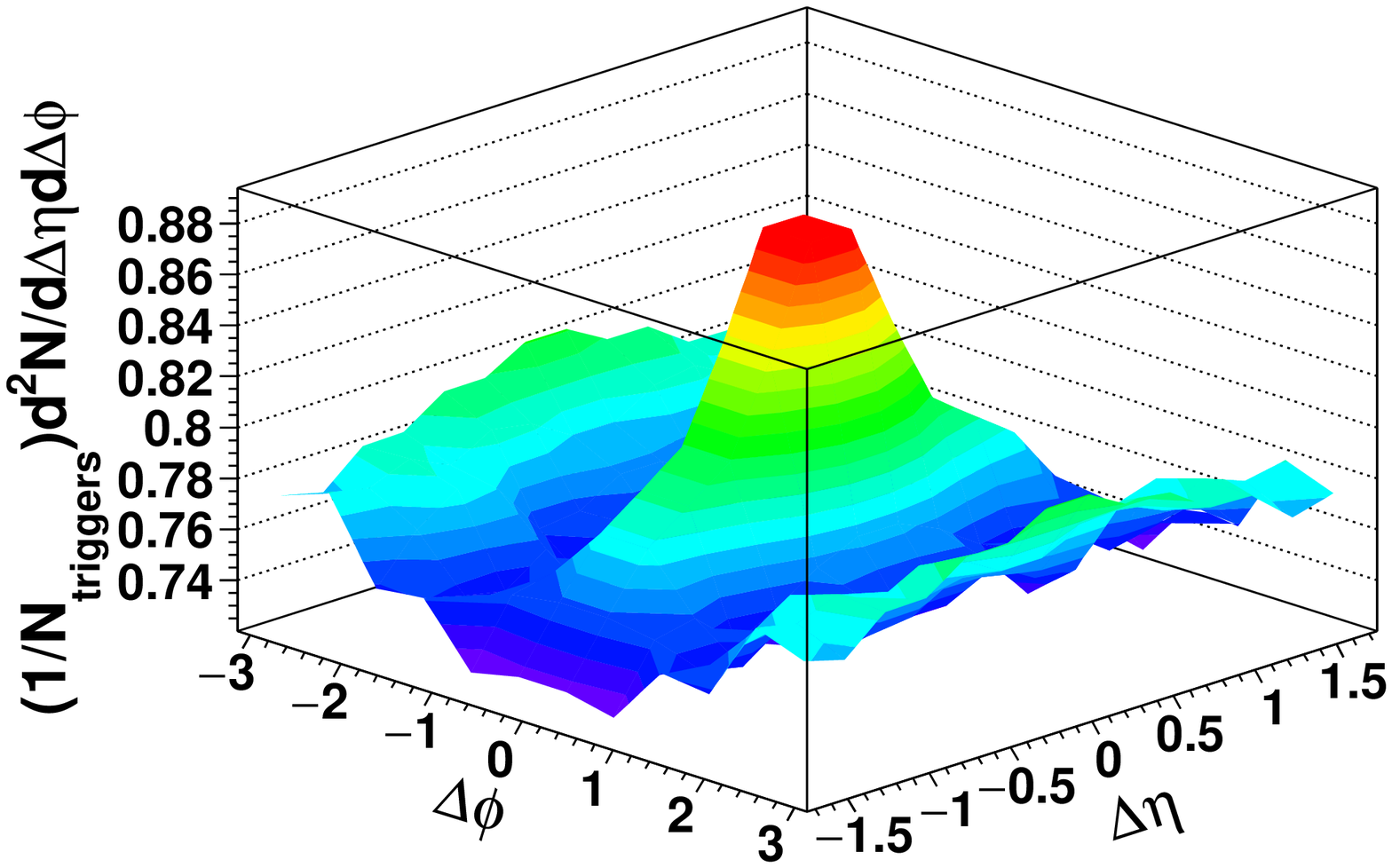}
}
}
\caption{(Color online.)  Corrected 2D \kaonh correlation function for \stdtrig and \stdassoc for 0-20\% \Cu.  The data have been reflected about \deta~=~0 and \dphi~=~0.}
\label{Fig:SampleCorr2D}
\end{figure}

\begin{figure}
\rotatebox{0}{\resizebox{8.5cm}{!}{
\includegraphics{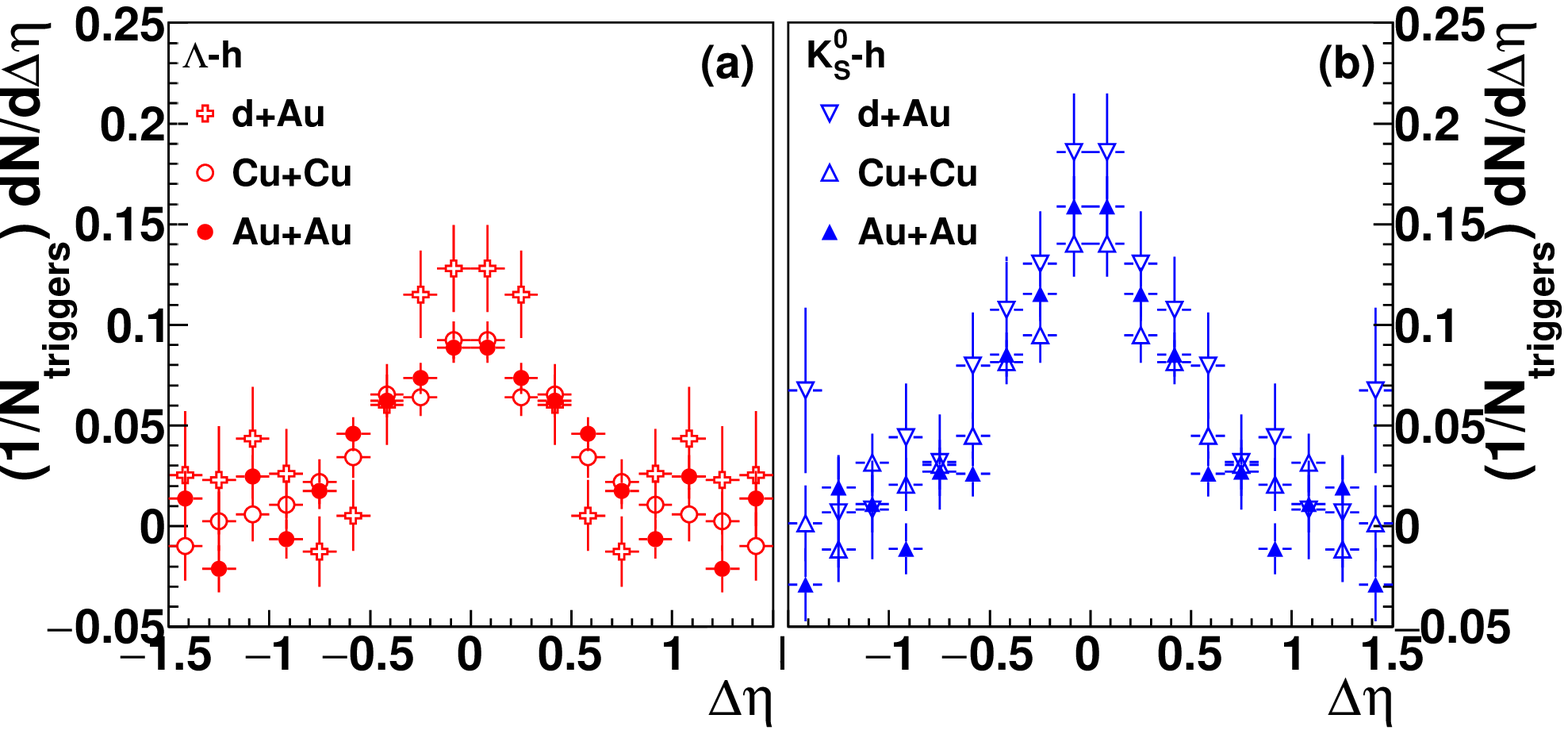}
}
}
\caption{(Color online.)  Corrected correlation functions $\frac{dN_{J}}{d\Deta}$ in $\mid$\dphi$\mid<$~0.78 for \stdtrig and \stdassoc for (a) \lamh and (b) \kaonh for minimum bias \dAu, 0-20\% \Cu, and 40-80\% \Au collisions at \sNNtwohundred  after background subtraction.  The data have been reflected about \deta~=~0.}
\label{Fig:SampleCorr}
\end{figure}

An example of a 2D correlation function after the corrections described above is shown in \Fref{Fig:SampleCorr2D}.  The notation and method used to extract the yield in this paper follow~\cite{Agakishiev:2011st,RidgePaper:2009qa}.
The \jlc is narrow in both \dphi and \deta and is contained within $|\Delta\phi|<0.78$ and $|\Delta\eta|<0.78$ for the kinematic cuts in \pttrig and \ptassoc used in this analysis.  The di-hadron correlation from \eref{eq:di-hadronDist} is projected onto the \deta axis:

\begin{eqnarray}
\label{eq:IDphi}
\left.\dNdEta\right|_{\Delta\phi_1,\Delta\phi_2}
\equiv\int_{\Delta\phi_1}^{\Delta\phi_2}d\Dphi\frac{d^2N}{d\Dphi{d}\Deta}.
\end{eqnarray}
\noindent  All other correlations, including those from \vtwo, \vthree, and higher order flow harmonics, are assumed to be independent of \deta within the $\eta$ acceptance of the analysis, consistent with ~\cite{Back:2004mh,phobosFlow2,RidgePaper:2009qa,Alver:2009id}.  We make the assumption that the $\eta$ dependence observed for \vthree measured using the two particle cumulant method~\cite{Adamczyk:2013waa} is entirely due to nonflow.  With these assumptions, both correlated and uncorrelated backgrounds such as flow are constant in \deta.  The \jlc can then be determined by:
\begin{eqnarray}
\label{eq:NJeta}
\frac{dN_{J}\left(\Deta\right)}{d\Deta}=
\left.\dNdEta\right|_{\Delta\phi_1,\Delta\phi_2}-b_{\Delta\eta}
\end{eqnarray}
\noindent
where $\bEta$ is a constant offset determined by fitting a constant background $b_{\Delta\eta}$ plus a Gaussian to $\frac{dN_{J}}{d\Deta}\left(\Deta\right)$.  Variations in the method for extracting the constant background, such as fitting a constant at large \deta, lead to differences in the yield smaller than the statistical uncertainty due to the background alone.  Nevertheless, a 2\% systematic uncertainty is applied to account for this.  This uncertainty is uncorrelated with the uncertainty on the efficiency for a total uncertainty of 5.5\% on all yields.  Examples of correlations are given in \Fref{Fig:SampleCorr}.  Where the track merging effect discussed below is negligible the yield from the fit and from bin counting are consistent.  When the dip due to track merging is negligible, the yield determined from fit is discarded to avoid any assumptions about the shape of the peak and instead we integrate \eref{eq:NJeta} over \deta using bin counting to determine the \jly $\Yjeta$:
\begin{align}
\Yjeta =  \int_{\Delta\eta_1}^{\Delta\eta_2} d\detano \; &
\frac{dN_{J}\left(\Deta\right)}{d\Deta}. \label{eq:Yjeta} 
\end{align}
\noindent    The choice of $\Delta\phi_1$, $\Delta\phi_2$, $\Delta\eta_1$, and $\Delta\eta_2$ is arbitrary.  For this analysis we choose $\Delta\phi_1=\Delta\eta_1$~=~$-0.78$ and $\Delta\phi_2=\Delta\eta_2$~=~0.78 in order to be consistent with previous studies and in order to include the majority of the peak~\cite{Agakishiev:2011st}. 

\subsection{Track merging correction}\label{section-method}
The track merging effect in unidentified particle (h) correlations discussed in~\cite{Agakishiev:2011st} is also present for \vzeroh and \hvzero correlations.  This effect leads to a loss of tracks at small \dphi and \deta due to overlap between the trigger and associated particle tracks and is manifested as a dip in the correlation function.  When one of the particles is a \vzero, this overlap is between one of the \vzero daughter particles and the unidentified particle.  The size of the dip due to track merging depends strongly on the relative momenta of the particle pair.  The effect is larger when the momentum difference of the two overlapping tracks is smaller.  For \vzeroh correlations, the typical associated particle momentum is approximately 1.5 \GeV.  Since the \kaon decay is symmetric, the track merging effect is greatest for \kaonh correlations with a trigger \kaon momentum of approximately 3 \GeV.  In a \lam decay, the proton daughter carries more of the \lam momentum than the pion daughter.  Therefore this effect is larger for \lam trigger particles with lower momenta.  Because track merging affects both signal and background particles and the signal sits on top of a large combinatorial background, the effect is larger for collisions with a higher charged track multiplicity.  Since the dip in \vzeroh and \hvzero correlations is the result of a \vzero daughter merged with an unidentified particle, the dip is wider in \dphi and \deta than in unidentified particle correlations.  

For identified \vzero associated particles in the kinematic range studied in this paper, there was no evidence for track merging.  A straightforward extension of the method in~\cite{Agakishiev:2011st} to \vzero trigger particles did not fully correct for track merging.  The residual effect was dependent on the helicity of the associated particle, demonstrating that this was a detector effect.  When the track merging dip is present, it is corrected by fitting a Gaussian to the peak, excluding the region impacted by track merging, and using the Gaussian fit to extract the yield.  The event mixing procedure described in~\cite{Agakishiev:2011st} was not applied to simplify the method since the yield would still need to be corrected using a fit to correct for the residual effect.

This correction is only necessary for the data points in \Fref{rjpttrig} specified below.  
To investigate the effect of using a fit where the peak is excluded from the fit region, we used a toy model where a Gaussian signal with a constant background was thrown with statistics comparable to the data with a residual track merging effect
When the peak is excluded from the fit for samples with high statistics, the yield is determined correctly from the fit.  For the low statistics samples comparable to the points with a residual track merging effect, the yield from the fit is usually within uncertainty of the true value but there is an average skew of about 13\% in the extracted yield.  A 13\% systematic uncertainty is added in quadrature to the statistical uncertainty on the yield from the fit so that these points can be compared to the other points.  
When the residual track merging effect is corrected by a fit, the track merging correction applied by the fit is approximately the same size as the statistical uncertainty on the yield.  We therefore conclude that when no dip is evident, the track merging effect is negligible compared to the statistical uncertainty on the yield.  

\subsection{Summary of systematic uncertainties}\label{sec-syserr}
Systematic uncertainties are summarized in \Tref{table-syserr}.
All data points have a 5\% systematic uncertainty due to the single track reconstruction efficiency and a 2\% systematic uncertainty due to the yield extraction method.  This is a total 5.5\% systematic uncertainty.  In addition, there is a 13\% systematic uncertainty due to the yield extraction for data points with residual track merging.  It is added in quadrature to the statistical uncertainty so that these data can be compared to data without residual track merging.  This uncertainty is only in the yields in \Fref{rjpttrig} listed below.

\begin{table}[b!]
\caption{Summary of systematic uncertainties due to the efficiency $\varepsilon$, yield extraction for all points, and yield extraction in the presence of a residual track merging effect.  The 13\% systematic uncertainty due to the yield extraction for data points with residual track merging is added in quadrature to the statistical uncertainty, which is on the order of 20-30\% for these data points.  This uncertainty is only in the yields in \Fref{rjpttrig} listed below.}
\label{tabevents}
\begin{tabular}{c c}\hline \hline
source  &  value (\%) \\ \hline \hline
$\varepsilon$    &  5\%  \\ 
yield extraction   &  2\% \\
yield with track merging (see caption)   &  13\% \\ \hline
total   &  5.5\% \\ \hline \hline
\end{tabular}\label{table-syserr}
\end{table}

\section{Results}\label{section-results}

\subsection{Charged particle-$V^0$ correlations}\label{hV0Corr}

Previous studies demonstrated that the \jlc in \hh correlations is nearly independent of collision system~\cite{Agakishiev:2011st,Abelev:2009ah,RidgePaper:2009qa}, with some indications of particle type dependence~\cite{Abdelwahab:2014cvd}, and that it is qualitatively described by PYTHIA~\cite{Agakishiev:2011st} at intermediate momenta.  This indicates that the \jlc is dominantly produced by fragmentation, even at intermediate momenta (2 $<$ \pT $<$ 6 \GeV) where recombination predicts significant modifications to hadronization.  The composition of the \jlc can be studied using correlations with identified associated particles.  For the analysis presented here, the size of \dAu data sample was limited and the \Au data set was limited by the presence of residual track merging.  Therefore it was only possible to determine the composition of the \jlc in \Cu collisions for a relatively large centrality range (0-60\%).  

\begin{figure}[ht!]
\begin{center}
\rotatebox{0}{\resizebox{8.0cm}{!}{
	\includegraphics{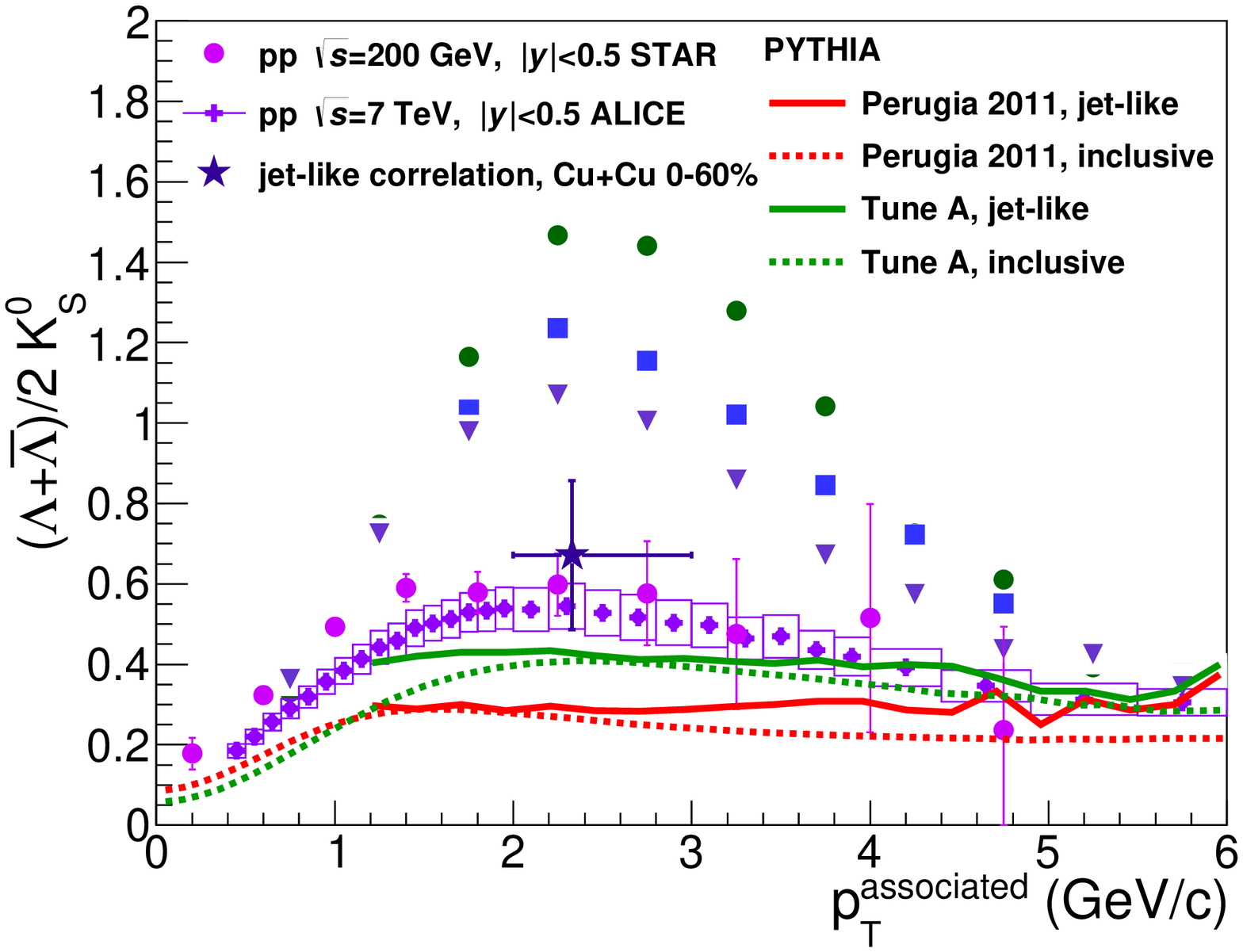}
}}
\caption{(Color online.)  $\Lambda$/$K^0_S$ ratio measured in the \jlc in 0-60\% \Cu collisions at \sNNtwohundred for \stdtrig and \assocrange{2.0}{3.0} along with this ratio obtained from inclusive $p_T$ spectra in \pp collisions.  Data are compared to calculations from PYTHIA~\cite{Sjostrand:2006za} using the Perugia 2011 tunes~\cite{Skands:2010ak} and Tune A~\cite{Field:2005sa}.}
\label{bmratio}
\end{center}
\end{figure}

These measurements are compared to inclusive baryon to meson ratios in \pp collisions from the STAR experiment~\cite{Abelev:2006cs} and the ALICE experiment~\cite{Aamodt:2011zza} and simulations of \pp collisions in PYTHIA~\cite{Sjostrand:2006za} using the Perugia 2011~\cite{Skands:2010ak} tune and Tune A~\cite{Field:2005sa} in~\Fref{bmratio}. The ratio in the \jlc in \Cu collisions is consistent with the inclusive particle ratios from \pp. This further supports earlier observations that the jet-like correlation in heavy-ion collisions is dominantly produced by the fragmentation process, which also governs the production of particles in \pp collisions at these momenta. It also implies that production of strange particles through recombination is not significant in the \jlc, even in \AplusA collisions, where the inclusive spectra show an enhancement of \lam production of up to a factor of three relative to the \kaon~\cite{Agakishiev:2011ar,Abelev:2013xaa}. 

The experimentally measured particle ratios in \pp collisions at \sqrts = 200 and 7000 GeV are consistent with each other.  However, they are not described well by PYTHIA. PYTHIA is able to match the light quark meson ($\pi$ and $\omega$) production~\cite{Aamodt:2010my,ALICE:2011ad}, but generally underestimates production of strange particles, especially strange baryons~\cite{Aamodt:2010my,ALICE:2011ad,Abelev:2006cs,Aamodt:2011zza}. Tune A has been adjusted to match low momentum h-h correlations~\cite{Field:2005sa}, while the Perugia 2011 tune has been tuned to match inclusive particle spectra better, including data from the LHC~\cite{Skands:2010ak}. The most recent MONASH tune~\cite{Skands:2014pea}, which is a variation of Tune A, had some success in capturing the inclusive strange meson yield at the LHC, but the \lam yield is still underestimated by a factor of 2. The discrepancy grows with the strange quark content of the baryon. 
Since \hvzero correlations are dominated by gluon and light quark jet fragmentation, PYTHIA underestimates the generation of strange quarks in those jets. This effect is enhanced in strange baryon production since the formation of an additional di-quark is required in PYTHIA. The probability of such a combination is significantly suppressed in PYTHIA, whereas the data seem to suggest that di-quark formation is not necessary to form strange baryons.  The discrepancy between PYTHIA and the data in~\Fref{bmratio} can therefore be attributed exclusively to the problems of describing strange baryon production in PYTHIA. 
On the other hand, strange particle triggered correlations, such as \kaonh and \lamh, originate predominantly from the fragmentation of strange quarks. It should be easier for PYTHIA to describe the production of strange particles from the fragmentation of strange quarks than light quarks and gluons. We therefore studied the \vzeroh correlations in more detail.

\subsection{Correlations with identified strange trigger particles}

\begin{figure}
\rotatebox{0}{\resizebox{8.0cm}{!}{\includegraphics{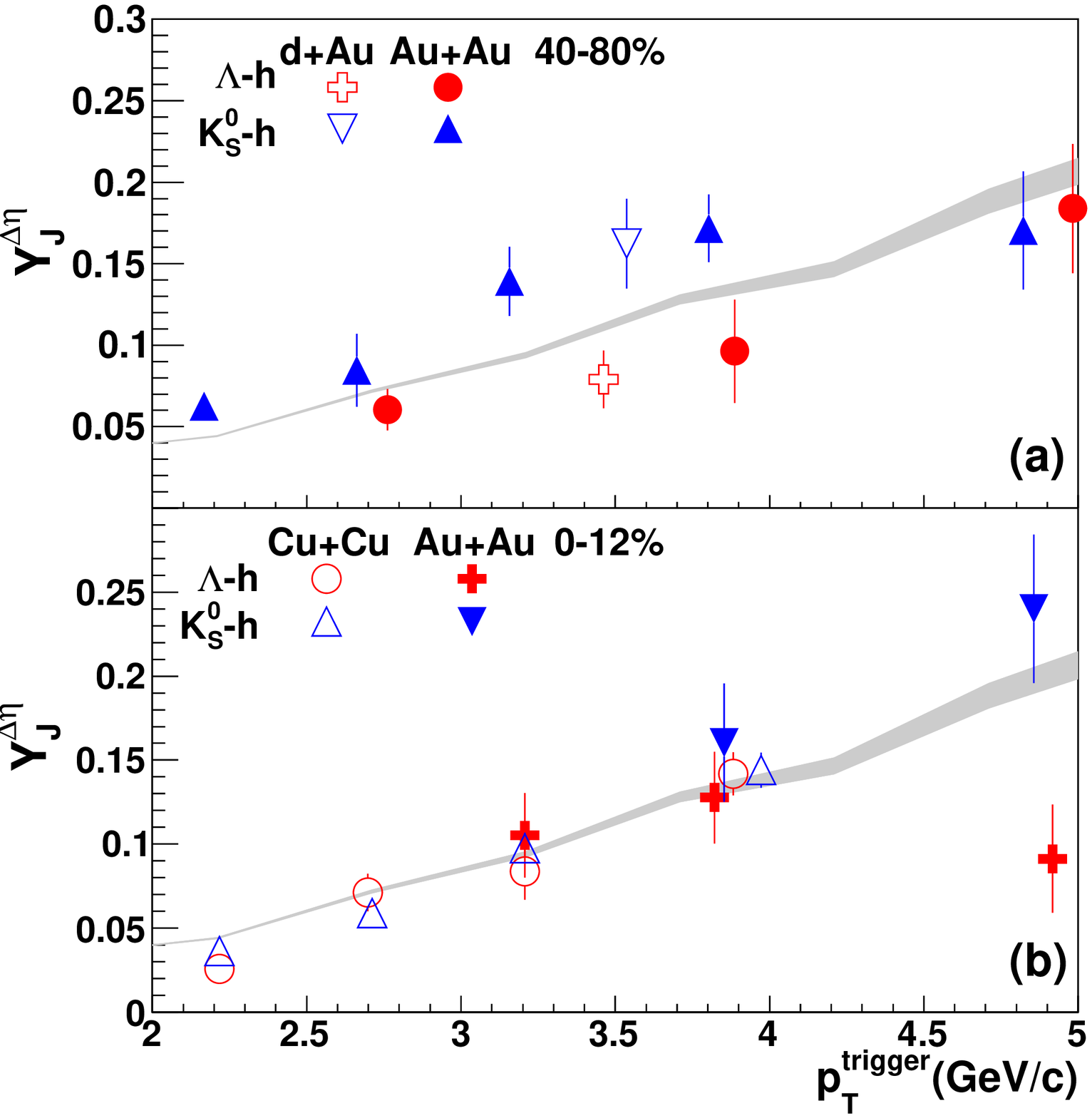}}}
\caption{(Color online.)  The jet-like yield in $\mid$\deta$\mid<$0.78 as a function of \pttrig for \kaonh and \lamh correlations for \stdassoc in (a) minimum bias  \dAu and 40-80\% \Au collisions at \sNNtwohundred and (b) 0-60\% \Cu and 0-12\% \Au collisions at \sNNtwohundred.  
For comparison \hh correlations~\cite{Agakishiev:2011st} from 40-80\% \Au collisions are shown as a band where the width represents the uncertainty.
Peripheral \Au points have been shifted in \pttrig for visibility.
\efficiencystatement}
\label{rjpttrig}
\end{figure}

The \jly as a function of \pttrig is shown in \Fref{rjpttrig} for \kaonh and \lamh correlations for \dAu, \Cu, and \Au collisions at \sNNtwohundred.  The data are tabulated in \Tref{trigpttable}.  
Due to residual track merging effects discussed in \Sref{section-method}, fits are used for \lamh correlations in some \pttrig ranges: in \Cu collisions, \trigrange{2.0}{3.0}; in 0-12\% Au+Au collisions, \trigrange{3.0}{4.5}; and in 40-80\% Au+Au collisions, \trigrange{2.0}{4.5}.
There is no significant difference in the yields between the collision systems, however, the data are not sensitive enough to distinguish the 20\% differences observed for identified pion triggers~\cite{Abdelwahab:2014cvd}.  No system dependence is observed for \hh correlations in~\cite{Agakishiev:2011st,Abdelwahab:2014cvd}.   This includes no significant difference between results from \Au collisions in 40-80\% and 0-12\% central collisions.
For this reason we only compare to \hh correlations from 40-80\% \Au collisions.

\begin{figure}
\rotatebox{0}{\resizebox{8.0cm}{!}{
	\includegraphics{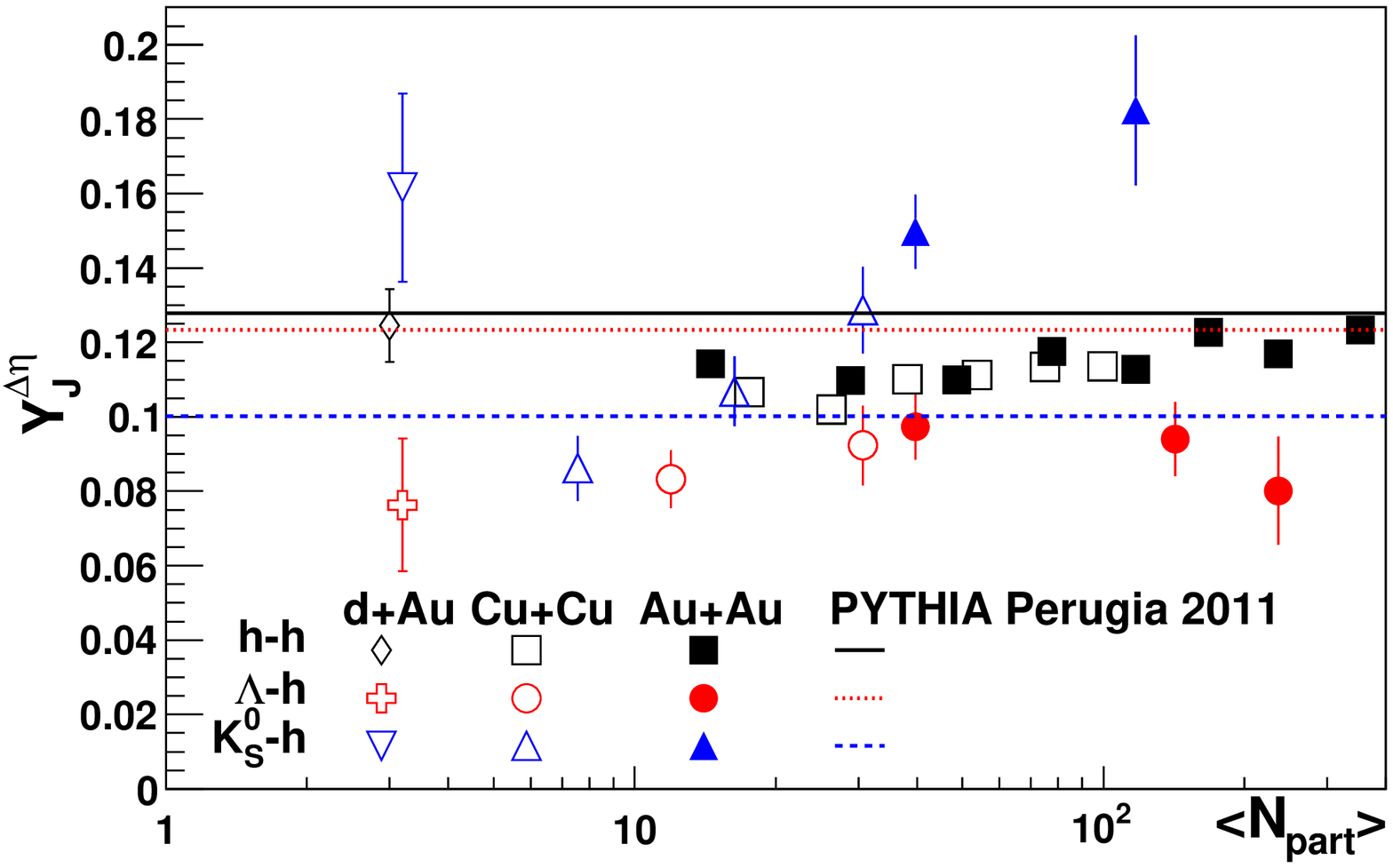}
}}
\caption{(Color online.)  Centrality dependence of the jet-like yield of \kaonh and \lamh correlations for \stdtrig and \stdassoc in \dAu, \Cu, and \Au collisions at \sNNtwohundred.  
The data are compared to PYTHIA~\cite{Sjostrand:2006za} calculations using the Perugia 2011 tune~\cite{Skands:2010ak}.  
\efficiencystatement} 
\label{rjcentr}
\end{figure}

Next the \jlys are studied as a function of collision centrality expressed in terms of number of participating nucleons (\npart) calculated from the Glauber model ~\cite{Miller:2007ri}. The extracted \jly as a function of \npart is shown in \Fref{rjcentr} for \hh~\cite{Agakishiev:2011st}, \kaonh, and \lamh correlations for \dAu, \Cu, and \Au collisions at \sNNtwohundred.  All yields are determined using bin counting.  
While there is no centrality dependence in the \jly of \hh correlations, there is a centrality dependence in the yields of the \kaonh correlations.
These data are compared to PYTHIA~\cite{Sjostrand:2006za} calculations from the Perugia 2011~\cite{Skands:2010ak} tune in \Fref{rjcentr}.  There is a hint of a particle species ordering, with the \jly from \kaonh correlations generally above that of the \jly from \hh correlations and the \jly from \lamh generally below that of the \hh correlations.  This is different from the particle type ordering observed in PYTHIA.

\begin{figure}
\rotatebox{0}{\resizebox{8.0cm}{!}{
	\includegraphics{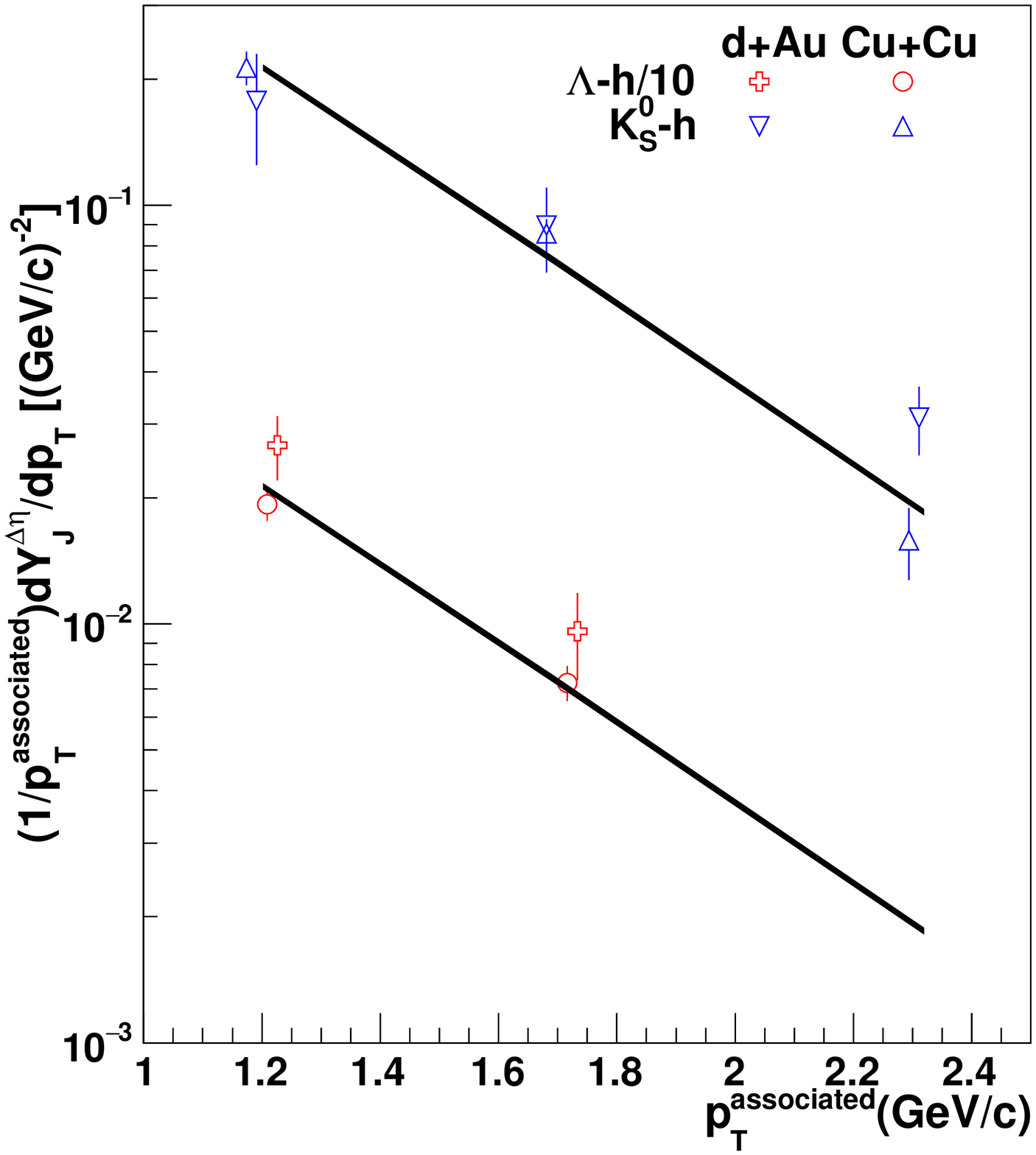}
}
}
\caption{(Color online.)  The \jly as a function of \ptassoc for \kaonh and \lamh correlations for \stdtrig in \dAu and 0-60\% \Cu collisions at \sNNtwohundred.  The data are compared to the \jly from \hh correlations~\cite{Agakishiev:2011st} from 40-80\% \Au collisions shown as a line.  Data are binned in \assocrange{1.0}{1.5}, \assocrange{1.5}{2.0}, and \assocrange{2.0}{3.0} and are plotted at the mean of the bin.  \efficiencystatement}
\label{rjptassoc}
\end{figure}

The \jly as a function of \ptassoc is shown in \Fref{rjptassoc} for \kaonh and \lamh correlations for \dAu and \Cu collisions at \sNNtwohundred.  All yields are determined using bin counting.  The \lamh and \kaonh correlations are only shown for \dAu and \Cu collisions since residual track merging made measurements in \Au collisions difficult.  Data are compared to the \jly from \hh correlations~\cite{Agakishiev:2011st}.  The trend is similar for \hh, \kaonh, and \lamh correlations, although the wide centrality bins required by low statistics may mask centrality dependencies such as those shown in \Fref{rjcentr}.

\begin{table}[b!]
\caption{The jet-like yield in $\mid$\deta$\mid<$0.78 as a function of \pttrig for \kaonh and \lamh correlations for \stdassoc in minimum bias \dAu, 0-60\% \Cu, and \Au collisions at \sNNtwohundred, as shown in \Fref{rjpttrig}.}
\label{trigpttable}
\vspace{0.5cm} 
\begin{tabular}{ c  c  c  c}
\hline \hline
Collision system, & \pttrig                    & {\kaonh} & {\lamh} \\  
  centrality      &  (GeV/$c$)             &    yield        &   yield \\  \hline \hline 
\dAu,        &  3.0-5.0 & 0.162 $\pm$ 0.028 & 0.079 $\pm$ 0.018 \\ 
0-95\%           &         &                   & \\ \hline
\Cu,        &  2.0-2.5   &   0.036 $\pm$ 0.004   & 0.026 $\pm$ 0.005 \\ 
0-60\%       &  2.5-3.0   &   0.059 $\pm$ 0.006   & 0.071 $\pm$ 0.007 \\
             &  3.0-3.5   &   0.098 $\pm$ 0.009   & 0.084 $\pm$ 0.017 \\
             &  3.5-5.0   &   0.144 $\pm$ 0.011   & 0.142 $\pm$ 0.013 \\ \hline

\Au,       & 2.0-2.5  &   0.063 $\pm$ 0.008   &  - \\ 
40-80\%      &  2.5-3.0   &   0.084 $\pm$ 0.023   & 0.061 $\pm$ 0.010 \\
             &  3.0-3.5   &   0.139 $\pm$ 0.022   &  - \\
             &  3.5-4.5   &   0.172 $\pm$ 0.021   & 0.096 $\pm$ 0.030 \\
             &  4.5-5.5   &   0.170 $\pm$ 0.037   & 0.184 $\pm$ 0.040 \\ \hline

\Au,       &  3.0-3.5   &    &   0.105 $\pm$ 0.021  \\ 
0-12\%       &  3.5-4.5   &  0.160 $\pm$ 0.036   &   0.128 $\pm$ 0.022  \\ 
              &  4.5-5.5   &  0.240 $\pm$ 0.045   & 0.091 $\pm$ 0.033  \\ \hline \hline

\end{tabular}
\end{table}

\section{Conclusions}\label{Conclusions}
Measurements of di-hadron correlations with identified strange associated particles demonstrated that the ratio of \lam to \kaon for the \jlc in \Cu collisions is comparable to that observed in \pp collisions.  This provides additional evidence that the \jlc is dominantly produced by fragmentation.  Measurements of di-hadron correlations with identified strange trigger particles show some centrality dependence, indicating that fragmentation functions or particle production mechanisms may be modified in heavy ion collisions.  These studies provide hints of possible mass ordering, although the measurements are not conclusive due to the statistical precision of the data.

These measurements provide motivation for future studies of strangeness production in jets.  Larger data sets and data from collisions at higher energies could provide more robust tests of the strangeness production mechanism.  Studies in \pp would be essential in order to search for modifications of strangeness production in jets in heavy ion collisions. 

\begin{acknowledgments}
\makeatletter{}We thank the RHIC Operations Group and RCF at BNL, the NERSC Center at LBNL, the KISTI Center in
Korea, and the Open Science Grid consortium for providing resources and support. This work was 
supported in part by the Office of Nuclear Physics within the U.S. DOE Office of Science,
the U.S. NSF, the Ministry of Education and Science of the Russian Federation, NSFC, CAS,
MoST and MoE of China, the National Research Foundation of Korea, NCKU (Taiwan), 
GA and MSMT of the Czech Republic, FIAS of Germany, DAE, DST, and UGC of India, the National
Science Centre of Poland, National Research Foundation, the Ministry of Science, Education and 
Sports of the Republic of Croatia, and RosAtom of Russia.
 
\end{acknowledgments}

\makeatletter{}

\end{document}